\documentclass[epj]{webofc}
\usepackage[utf8]{inputenc}
\usepackage[varg]{txfonts}   % Web of Conferences font
\usepackage{booktabs}
\usepackage{xcolor}
\definecolor{darkred}{rgb}{0.4,0.0,0.0}
\definecolor{darkgreen}{rgb}{0.0,0.4,0.0}
\definecolor{darkblue}{rgb}{0.0,0.0,0.4}
\usepackage[bookmarks,linktocpage,colorlinks,
    linkcolor = darkred,
    urlcolor  = darkblue,
    citecolor = darkgreen]{hyperref}
\usepackage{subfigure}
\wocname{EPJ Web of Conferences}
\woctitle{Lattice2017}
\begin{document}
%%%%%%%%%%%%%%%%%%%%%%%%%%%%%%%%%%%%%%%%%%%%%%%%%%%%%%%%%%%%%%%%%%%%%%%%%%%%%
%
\selectlanguage{english}
%----------------------------------------------------------------------------
\title{%
Update on SU(2) gauge theory with $N_F=2$ fundamental flavours.
}
%----------------------------------------------------------------------------
\author{%
\firstname{Vincent} \lastname{Drach}\inst{1},
\firstname{Tadeusz} \lastname{Janowski}\inst{2}\fnsep\thanks{Speaker, \email{T.Janowski@ed.ac.uk}}\and
\firstname{Claudio} \lastname{Pica}\inst{2}
% etc.
}
%----------------------------------------------------------------------------
\institute{%
University of Plymouth, Drake Circus, Plymouth, Devon, PL4 8AA, United Kingdom
\and
CP$\mathit{\,^3}$-Origins, University of Southern Denmark, Campusvej 55, DK-5230 Odense M, Denmark
%   \and
% School of Physics and Astronomy, University of Edinburgh, James Clerk Maxwell Building, Peter Guthrie Tait Road, Edinburgh, EH9 3FD, United Kingdom
}
%----------------------------------------------------------------------------
\abstract{We present a non perturbative study of SU(2) gauge theory with two fundamental Dirac flavours. This theory provides a minimal template which is ideal for a wide class of Standard Model extensions featuring novel strong dynamics, such as a minimal realization of composite Higgs models. We present an update on the status of the meson spectrum and decay constants based on increased statistics on our existing ensembles and the inclusion of new ensembles with lighter pion masses, resulting in a more reliable chiral extrapolation. \\[.1cm]
{\footnotesize  \it Preprint: CP3-Origins-2017-048 DNRF90  }
}
%----------------------------------------------------------------------------
\maketitle
%----------------------------------------------------------------------------
\section{Introduction}

The discovery of the Higgs boson in 2012 was a major breakthrough which confirmed the missing piece of the Standard Model (SM). However the SM Higgs sector is not free from theoretical issues. Maybe the most prominent of them is the naturalness problem, which can be formulated as follows. 
If we assume that the Standard Model is valid up to energy scale $\Lambda$ (which can be as high as the Planck scale), then the Higgs boson pole mass will be given by $m_H^2 = m_R^2(\Lambda) - \Sigma(\Lambda)$, where $m_R$ is a renormalised mass, which depends on the unknown UV details of the theory, and $\Sigma$ is the self-energy of the Higgs boson which is proportional to the square of the cutoff $\Sigma \propto \Lambda^2$.
This means that if SM is valid up to the Planck scale $10^{19}$ GeV then in order to achieve the experimentally measured value of 126 GeV, we need large cancellations between two unrelated quantities, which seems highly unnatural.

An elegant solution for this problem is to make the Higgs boson a composite particle, which would naturally have a mass of the dynamically generated compositeness scale. 
There are two main classes of such models, namely technicolor and composite (Goldstone) Higgs models.

In technicolor models \cite{Weinberg:1975gm, Susskind:1978ms} the generation of a techniquark condensate breaks the electroweak symmetry giving masses to W and Z bosons. The physical Higgs boson is then identified with the lightest scalar resonance, analogous to the scalar $\sigma$ resonance in QCD.
Realistic (extended) Technicolor models are difficult to achieve due to two major difficulties, namely mass generation of quarks and consistency with electroweak precision observables.
The quark masses can be generated, at the effective level, via 4-fermion couplings to SM fermions of the form $\frac{C_{QQqq}}{\Lambda_{ETC}^2}\bar q q \bar Q Q$, where $q$ denote SM quarks and $Q$ denote techniquarks, which can arise at the $\Lambda_{ETC}$ scale from a diagrams involving some heavy vector mediator.
Such extended technicolor interactions would generically also produce interactions among four SM fermions, such as $\frac{C_{\bar s d \bar s d}}{\Lambda_{ETC}^2}\bar s d \bar s d$. 
These in turn would generate flavour changing neutral currents, which are severely constrained by the measurement of the $K_L-K_S$ mass difference. To escape this experimental constraint one is then forced to push $\Lambda_{ETC}$ up to $\sim$1000 TeV, and assume the ETC interactions are CP-conserving.
Such a high value of $\Lambda_{ETC}$ leads to large suppression of quark masses, so that other mechanisms are needed to obtain realistic models.
A second issue arises from the contribution of the new strong sector to the electroweak precision observables, especially the Peskin-Takeuchi S parameter\cite{Peskin:1990zt}. 
In fact the presence of new techniquarks affects the vacuum polarisations of electroweak vector bosons and shifts the value of the S-parameter from the reference of $0$ for the SM\footnote{The T parameter can be made to vanish by requiring a global SU(2)$_L$ $\times$ SU(2)$_R$ custodial symmetry.}. 
The S-parameter therefore provides an important constraint on the allowed new physics and it is typically too large in naive technicolor models, such as scaled-up QCD.

A possible resolution of these problems is to invoke a novel type of strong dynamics, known as walking dynamics. 
Walking technicolor results in an enhancement of the quark masses respect to the naive case by a factor $(\Lambda_{ETC}/\Lambda_{TC})^\gamma$, where $\Lambda_{TC}$ is the scale at which the technicolor condensate forms ($\simeq$ the electroweak scale) and $\gamma$ is the anomalous dimension of the four-fermion operator responsible for the generation of SM quark masses. A large $\gamma \approx 1$ can produce an enhancement of quark masses sufficient to reproduce masses up to bottom quark mass. 

The second class of composite models goes under the name of composite (Goldstone) Higgs models \cite{Kaplan:1983fs, Kaplan:1983sm}. 
Similarly to technicolor, such models feature a new strong sector with hyperquarks featuring an enlarged global flavor symmetry.
In contrast to technicolor models, the electroweak symmetry is not broken by the formation of the hyperquark condensate at $\Lambda_{TC}$. The global flavor symmetry breaking of the new strong sector generates a number of Goldstone bosons, four of which take the role of the SM Higgs doublet. 
Interactions with SM fields give rise to an effective potential for these Goldstone bosons which results in electroweak symmetry breaking.
The amount of breaking is parametrised by the vacuum (mis)alignment angle $\theta$, which relates the electroweak scale $v=246$ GeV with TC scale given by the value of the decay constant $F$ via the relation $v = F\sin \theta$. 
The limit $\theta = \pi/2$ corresponds to the technicolor limit discussed above.
By tuning the misalignment angle to be small $\simeq 0.1$, one can achieve a little hierarchy between the TC scale and the electroweak scale thus explaining a light Higgs.
Similarly, the S-parameter is proportional to $\sin\theta$, so that experimental contraints can be resolved.

The concrete realization of composite Goldstone Higgs model considered here is SU(2) with 2 flavours of hyperquarks in fundamental representation \cite{Cacciapaglia:2014uja}. 
This model has already been studied on a lattice \cite{Arthur:2016dir, Hietanen:2014xca} and we present here an update of these studies. 
We remind the reader that the fermion representation is pseudo-real and, as a consequence, we can combine the hyperquark fields into the following multiplet:

  \begin{equation}
    Q = \left( \begin{array}{c}u_L\\d_L\\ -i\sigma^2 C \bar u_R^T\\ -i\sigma^2 C \bar d_R^T\end{array} \right)\, ,
  \end{equation}
  so that $Q$ transforms under a SU(4) flavour group.
  This global SU(4) symmetry is broken spontaneously by a fermion condensate $\Sigma^{ab} = \langle Q^a (i\sigma^2_c) C Q^b \rangle$ to the invariant subgroup:
  \begin{equation}
    U^T \Sigma U = \Sigma \quad U \in \mathrm{Sp(4)} \sim \mathrm{SO(5)}.
  \end{equation}
  The resulting Sp(4) $\sim$ SO(5) group contains the custodial symmetry group SO(4) $\sim$ SU(2)$_L$ $\times$ SU(2)$_R$ as a subgroup.
  The model has five Goldstone bosons, which transform under (2,2) and (1,1) representations of the custodial group, the former four can be associated with the SM Higgs doublet, while latter  state is a new SM-neutral yet-undiscovered scalar state\footnote{Such particle cannot be a dark matter candidate as it would presumably not be long-lived~\cite{Arbey:2015exa}.}.

\section{SU(2) spectrum}

In this section we report for the first time the theoretical predictions about the spectrum of the SU(2) model with 2 fundamental flavours of quarks.
The same arguments apply for any pseudo-real representation with two flavours of quarks.
As we noticed in the introduction, in the pseudo-real representation the flavour transformations will mix left- and right-chiral Weyl fields.
The same holds true for parity transformation defined as:
\begin{align}
  u_L(x) &\to \eta u_R(-x) \, ,\\
  u_R(x) &\to \eta u_L(-x) \, ,
\end{align}
where $\eta$ is an arbitrary phase.
It can be shown that the quark fields transform under parity as
\begin{align}
  Q(x) \to 
  \left(
  \begin{array}{cc}
    0 & \eta 1_2 \\
    \eta^* 1_2 & 0
  \end{array}
  \right)
  (i\sigma^2_c)(i\sigma^2_s) Q^*(x^P) \, ,
\end{align}
while the flavour transformation is as usual
\begin{equation}
  Q(x) \to M Q(x) \, ,
\end{equation}
where $M$ is a flavour rotation matrix.
Requiring that parity commutes with flavour symmetry (i.e. equating flavour followed by parity to parity followed by flavour) gives
\begin{equation}
  \left(
  \begin{array}{cc}
    0 & \eta 1_2 \\
    \eta^* 1_2 & 0
  \end{array}
  \right) M^* = M 
  \left(
  \begin{array}{cc}
    0 & \eta 1_2 \\
    \eta^* 1_2 & 0
  \end{array}
  \right)\, .
\end{equation} 

This relation does not hold for an arbitrary SU(4) matrix. On the other hand, if we choose $M$ such that it leaves a vacuum condensate $\Sigma$ invariant, i.e.
\begin{align} 
  M \Sigma M^T &= \Sigma \, ,\\
  M^\dag M &= 1 \, ,
\end{align}
we then have
\begin{equation}
  \left(
  \begin{array}{cc}
    0 & \eta 1_2 \\
    \eta^* 1_2 & 0
  \end{array}
  \right) \Sigma^{-1} M = M 
  \left(
  \begin{array}{cc}
    0 & \eta 1_2 \\
    \eta^* 1_2 & 0
  \end{array}
  \right)\Sigma^{-1} \, .
\end{equation}
Because $M$ is in an irreducible representation of Sp(2N), from Schur's lemma, any matrix that commutes with it must be proportional to the identity.
We therefore have
\begin{equation}
  \left(
  \begin{array}{cc}
    0 & \eta 1_2 \\
    \eta^* 1_2 & 0
  \end{array}
  \right) \propto \Sigma \, .
\end{equation} 

When $\Sigma = E \equiv
  \left(
  \begin{array}{cc}
    0 & 1_2 \\
    - 1_2 & 0
  \end{array}
  \right) $
this condition is satisfied if $\eta = -\eta^*$, i.e. $\eta = \pm i$.
We remark that for any other choice of the condensate, this definition of parity is not sufficient and we must instead use a combination of parity and flavour rotation.

Now that we have defined parity in a way which is consistent with flavour transformations, we will now classify mesonic states.
As (Hyper)quarks are in the fundamental representation of Sp(4), we can classify all the possible mesons made of two quarks:
\begin{equation}
  \mathbf 4 \otimes \mathbf 4 = \mathbf{10} \oplus \mathbf 5 \oplus \mathbf 1 \, ,
\end{equation}
i.e. mesons can be either in a 10-dimensional, 5-dimensional or a singlet representation of the flavour group.

As pions are in the 5-dimensional representation, states consisting of 2 or 3 pions can also be classified by symmetry:
\begin{align}
  \mathbf 5 \otimes \mathbf 5 &= \mathbf{14} \oplus \mathbf{10} \oplus \mathbf 1\, ,\\
  \mathbf 5 \otimes \mathbf 5 \otimes \mathbf 5 &= \mathbf{35} \oplus \mathbf{35}\oplus \mathbf{30}\oplus \mathbf{10}\oplus \mathbf 5\oplus \mathbf 5\oplus \mathbf 5\, .
\end{align}

In conclusion, decuplet ({\bf 10}) mesons can decay to either 2 or 3 pions, pentuplet ({\bf 5}) can only decay to 3 pions, while singlets can only decay to 2 pions.

To construct interpolating operators with the right quantum numbers, we consider the representations of flavour group together with the spin representations. The relevant flavour irreps are given by:
\begin{itemize}
  \item symmetric (dim 10): $\varphi^a \chi^b + \varphi^b \chi^a$ \, ,
  \item antisymmetric, ``E-traceless'' (dim 5): $\varphi^a \chi^b - \varphi^b \chi^a - \frac{E^{ab}}{2} \varphi^c E_{cd} \chi^d$\, ,
  \item ``E-trace'' $\varphi^a E_{ab} \chi^b$\, ,
\end{itemize}
where the tensor $T^{ab}$ is understood to be ``E-traceless'' if $E_{ab}T^{ab} = 0$.
For the spin we have:
\begin{itemize}
  \item Scalar: $Q^a (i\sigma^2_c)(i\sigma^2_s) Q^b \pm \tilde Q^a (i\sigma^2_c)(i\sigma^2_s) \tilde Q^b$ (with + sign corresponding to parity odd operator)\, ,
  \item Vector: $\tilde Q^a (i\sigma^2_c)(i\sigma^2_s) \bar \sigma^\mu Q^b $ \, ,
  \item Tensor: $Q^a (i\sigma^2_c)(i\sigma^2_s) \sigma^{\mu\nu} Q^b \pm \tilde Q^a (i\sigma^2_c)(i\sigma^2_s) \bar \sigma^{\mu\nu} \tilde Q^b$\, ,
\end{itemize}
where $\tilde Q = (i\sigma^2_c)(i \sigma^2_s) E Q^*$.

We report in Table \ref{tab:spectrum} our results for the mesonic states and interpolating operators.

\begin{table}
\begin{center}
\begin{tabular}{c|c|c|c}
  Flavour & $J^P$ & operator & decay products \\
  \hline
  $\mathbf{5}$ & $0^-$ & $\bar U \gamma^5 D$ & stable\\
  \hline
  $\mathbf{1}$ & $0^-$ & $\bar U \gamma^5 U + \bar D \gamma^5 D$ & stable\\
  $\mathbf{1}$ & $0^+$ & $\bar U U + \bar D D$ & 2 $\pi$ s-wave\\
  $\mathbf{5}$ & $0^+$ & $\bar U D$ & stable\\
  $\mathbf{1}$ & $1^+$ & $\bar U \gamma^i \gamma^5 U + \bar D \gamma^i \gamma^5 D$ & stable\\
  $\mathbf{10}$ & $1^-$ & $\bar U \gamma^i D$ & 2$\pi$ p-wave, 3$\pi$ p-wave\\
  $\mathbf{5}$ & $1^+$ & $\bar U \gamma^i \gamma^5 D$ & 3 $\pi$ p-wave\\
  $\mathbf{10}$ & $1^+$ & $\bar U \gamma^0\gamma^5\gamma^i D$ & 3$\pi$ p-wave
\end{tabular}
\end{center}
\caption{Expected states in the spectrum of SU(2) theory together with examples of interpolating operators that overlap with them and the expected decay products.}
\label{tab:spectrum}
\end{table}
\section{Lattice setup}
We summarize here the lattice setup used in our simulations, more details can be found in \cite{Arthur:2016dir,Arthur:2016ozw}. We use Wilson SU(2) gauge action with unimproved Wilson fermion.
The spectrum is obtained from two-point functions of the appropriate interpolating operators reported above, estimated by using stochastic quark sources with $Z_2\times Z_2$ noise at the source \cite{Boyle:2008rh}.
\begin{table}
  \begin{center}
  \begin{tabular}{c|c|l}
    beta & volume & $-a m_0$\\
    \hline
    1.8 & $16^3 \times 32$ & 1.0, 1.089, 1.12, 1.14, 1.15\\
    1.8 & $24^3 \times 32$ & 1.155, 1.157 \\
    2.0 & $16^3 \times 32$ & 0.85, 0.9, 0.94, 0.945 \\
    2.0 & $32^3 \times 32$ & 0.947, 0.949, 0.952, 0.956, $0.957^1$, $0.959^1$ \\
    2.0 & $32^3 \times 64$ & $0.958^2$, $0.9585^3$ \\
    2.2 & $16^3 \times 32$ & 0.6, 0.65, 0.68, 0.7 \\
    2.2 & $32^3 \times 32$ & 0.72, 0.735, 0.75 \\
    2.2 & $48^3 \times 48$ & $0.76^1$, $0.763^3$\\
    2.3 & $32^3 \times 32$ & 0.575, 0.6, 0.625, 0.65, $0.675^1$\\
    2.3 & $48^3 \times 48$ & $0.68^3$, $0.685^2$
  \end{tabular}
\end{center}
\caption{List of ensembles used in this run. Numbers denotes the ensembles which were updated since \cite{Arthur:2016dir}. 1 - increased statistics, 2 - increased volume, 3 - new quark mass. }
\end{table}
We use Wilson flow \cite{Luscher:2011bx} quantity $w_0$ to set the scale, defined via smooth fields defined at some flow time $t$ by:
  \begin{equation}
    \partial_t B_\mu = D_\nu G_{\nu\mu}\, ,
  \end{equation}
with $\left. B_\mu \right|_{t=0} = A_\mu$ and $G_{\mu\nu} = \partial_\mu B_\nu - \partial_\nu B_\mu + [B_\mu,B_\nu]$. To set the lattice spacing we consider the following quantity:
  \begin{equation}
    W(t) = t \frac{d}{d t} \langle t^2 E(t) \rangle\, ,
  \end{equation}
where $E=\frac14 G^a_{\mu\nu}G^a_{\mu\nu}$ is the discretized gauge action for the smooth fields, and for each simulation we define $w_0$ by the condition:
  \begin{equation}
    W(w_0^2/a^2) = 1.
  \end{equation}
  The lattice spacing is obtained from a chiral extrapolation via a fit of the form:
  \begin{equation}
    w_0(m_\pi) = w_0^\chi (1 + Ay^2 + By^4 \log y^2)\, ,
  \end{equation}
  with $y \equiv w_0(m_{PCAC})m_\pi$.
  The updated values of $w_0$ are summarised in Table \ref{tab:w0}.
  
  \begin{table}
    \begin{center}
      \begin{tabular}{c|cccc}
	$\beta$ & 1.8 & 2.0 & 2.2 & 2.3\\
	\hline
	$w_0/a$ & 2.068(26) & 2.711(17) & 4.318(58) & 6.369(514)
      \end{tabular}
    \end{center}
    \caption{The updated value of $w_0^\chi/a$ in the chiral limit}
    \label{tab:w0}
  \end{table}

  Finally, we will present below some quantities, such as $m_{PCAC}$ and decay constants, which require the knowledge of appropriate renormalisation constants.  We use the RI'-MOM scheme \cite{Sachrajda:2004mi} as described in \cite{Arthur:2016dir}.
\section{Results}
\subsection{$\chi_{PT}$ parameters}
Low energy constants of the chiral Lagrangian are obtained from a combined chiral and continuum limit as described in \cite{Arthur:2016dir}. The functional form we use is given by:
  \begin{align}
    f_{\pi} &= F(1 + C_1 m_\pi^2 \log m_\pi^2 + C_2 m_\pi^2 + C_3/w_0 + C_4 m_\pi^2/w_0)\, \\
    m_\pi^2/m_f &= 2B(1 + C_1' m_\pi^2 \log m_\pi^2 + C_2' m_\pi^2 + C_3'/w_0 + C_4' m_\pi^2/w_0)\, .
  \end{align}

  % \begin{figure}
  %   \begin{center}
  %     \includegraphics[width=0.45\textwidth]{mps2mpcac.pdf}
  %     \includegraphics[width=0.45\linewidth]{fps.pdf}
  %   \end{center}
  %   \caption{Fit to $\chi PT$ parameters using $f_{PS}$ or $m_\pi^2/m_{PCAC}$ as a function of $m_\pi^2$.}
  %   \label{fig:chiral_all}
  % \end{figure}

\begin{figure}
    \begin{center}
    \includegraphics[width=0.45\linewidth]{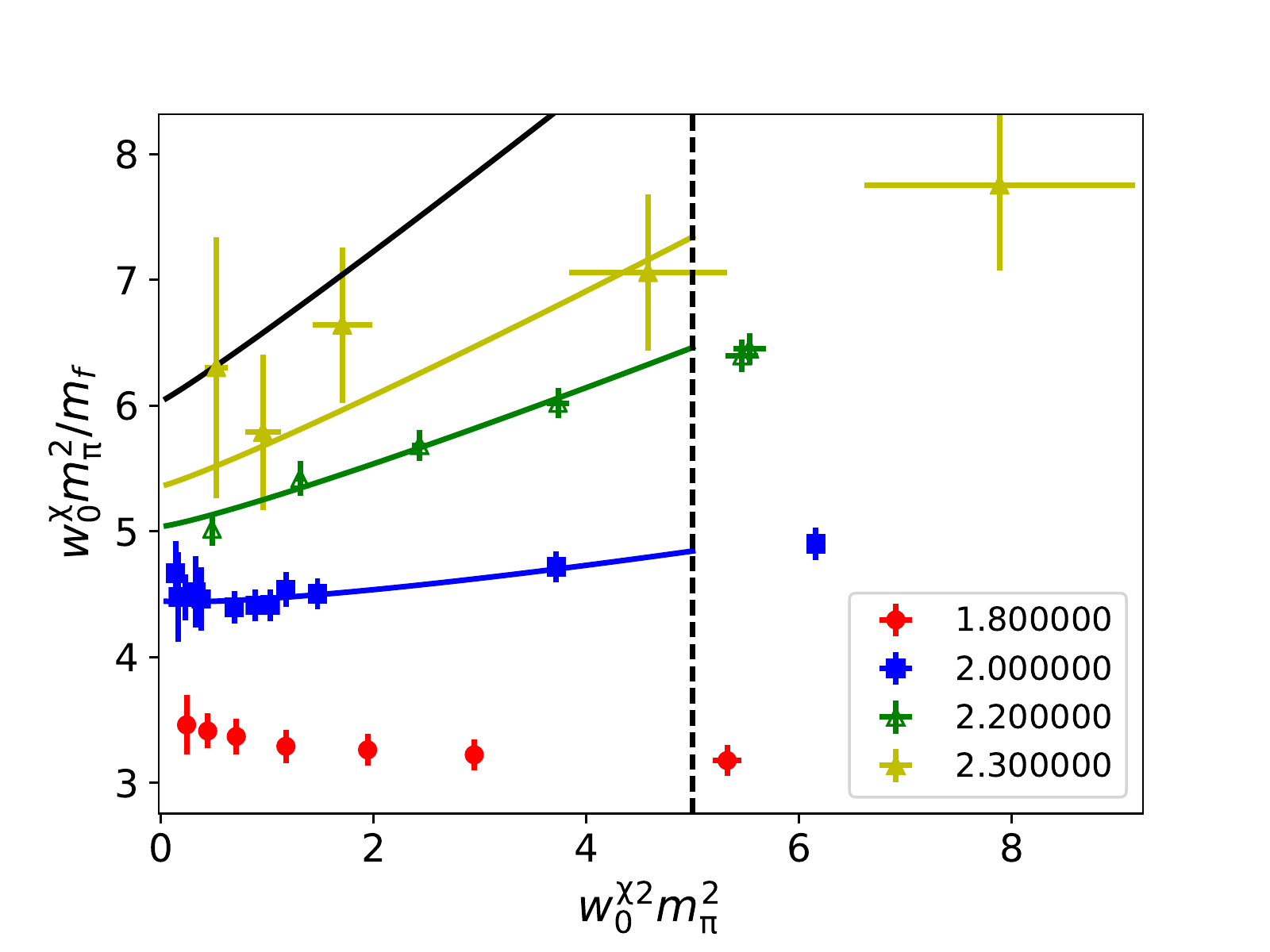}
    \includegraphics[width=0.45\linewidth]{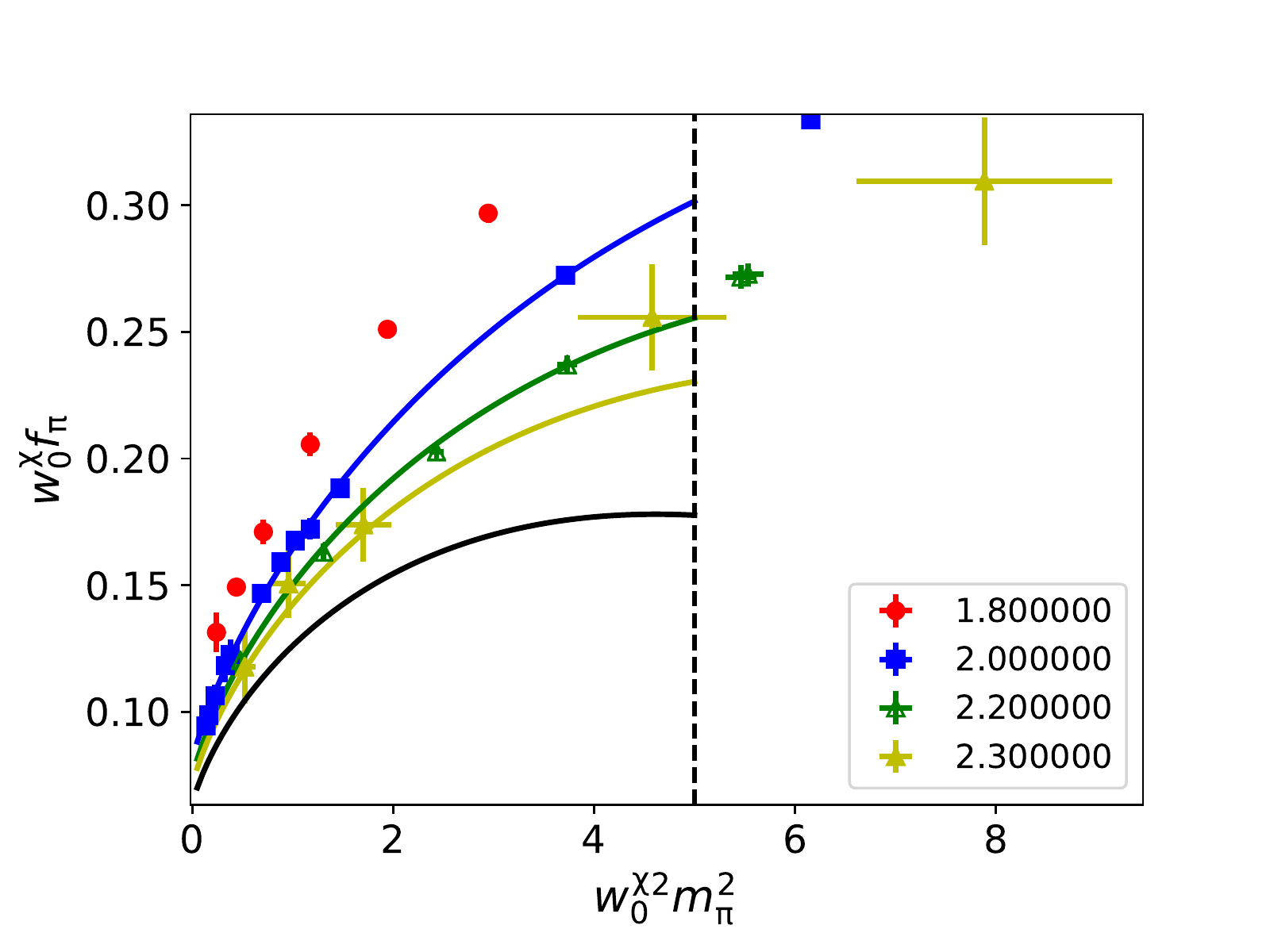}
  \end{center}
  \caption{Combined chiral and continuum fit of $m_\pi^2/m_{PCAC}$ (left) and $f_{PS}$ (right) as function of $m_\pi^2$. The data on the coarsest lattice spacing at $\beta = 1.8$ have been excluded.}
  \label{fig:chiral_no18}
\end{figure}

As previously reported in \cite{Arthur:2016dir}, our coarset lattice ensemble has large discretization effects for $f_\pi$ which does not fit well the formula above producing a $\chi^2/dof = 2.01$ with the updated dataset used here.
Excluding the $\beta = 1.8$ data from the fit yields a good fit, shown in Figure \ref{fig:chiral_no18}.

The values for the best fit parameters are
  $F = 0.0632(55)$ with $\chi^2/dof = 0.456 $ and
  $B = 3.01(11)$  with $\chi^2/dof = 0.328$ .

\subsection{Meson masses}
We also study the vector and the axial mesons, the latter in both representations of the flavour group. Following the same procedure as in \cite{Arthur:2016dir}, the meson masses are extracted from a combined chiral and continuum extrapolation via the phenomenological formula:
\begin{equation}
    w_0^\chi m = w_0^\chi m^\chi + A (w_0^\chi m_\pi)^2 + B (w_0^\chi m_\pi)^4 + C \frac{a}{w_0}\, .
\end{equation}
\begin{figure}
  \begin{center}
    \includegraphics[width=0.65\linewidth]{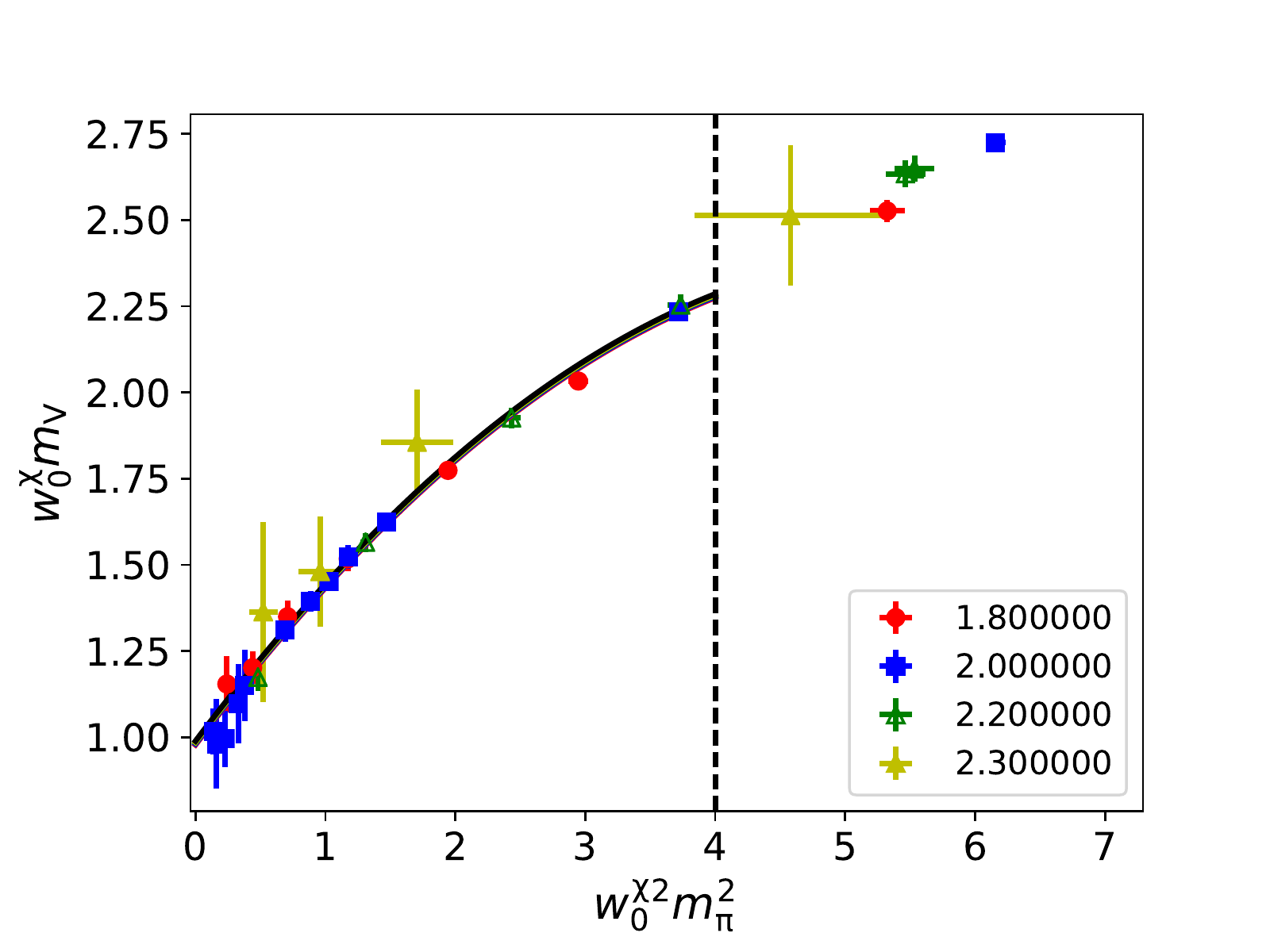}
  \end{center}
  \caption{Chiral and continuum extrapolation of vector meson mass.}
  \label{fig:mv}
\end{figure}
The fit to the vector meson mass is shown in Figure~\ref{fig:mv}.
The best fit value of the vector mass is
  $w_0^\chi m_V^\chi = 0.989(27)$ with $\chi^2/dof = 0.378$. 
We find that the cutoff effects are within our statistical errors for this quantity and that the vector meson mass is very close to 1 in the units of $w_0^\chi$.

 Fits for the axial meson masses are shown in Figure \ref{fig:ma}. The extracted values of the masses in the chiral and continuum limit are:
 \begin{figure}
  \begin{center}
    \includegraphics[width=0.45\linewidth]{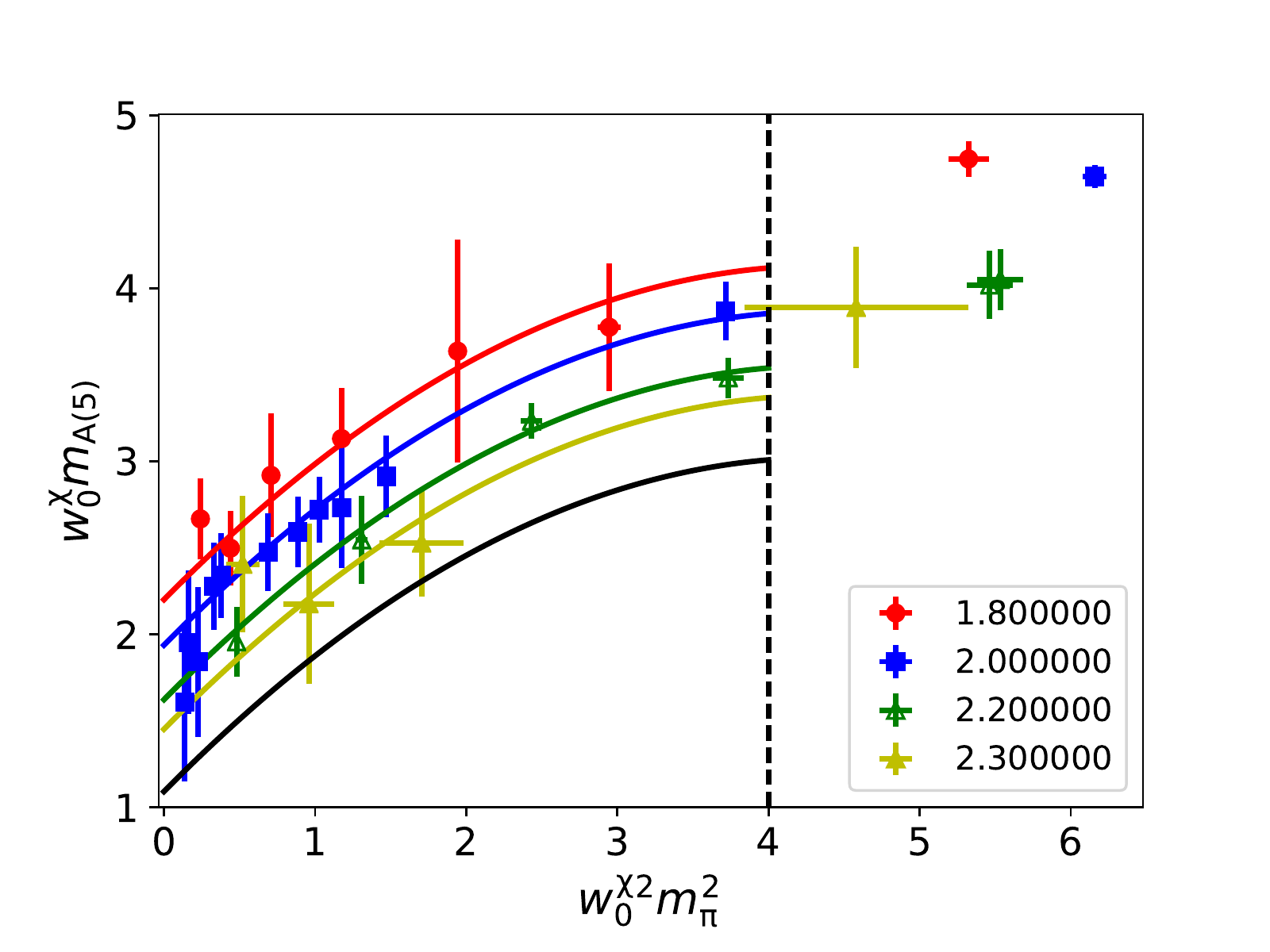}
    \includegraphics[width=0.45\linewidth]{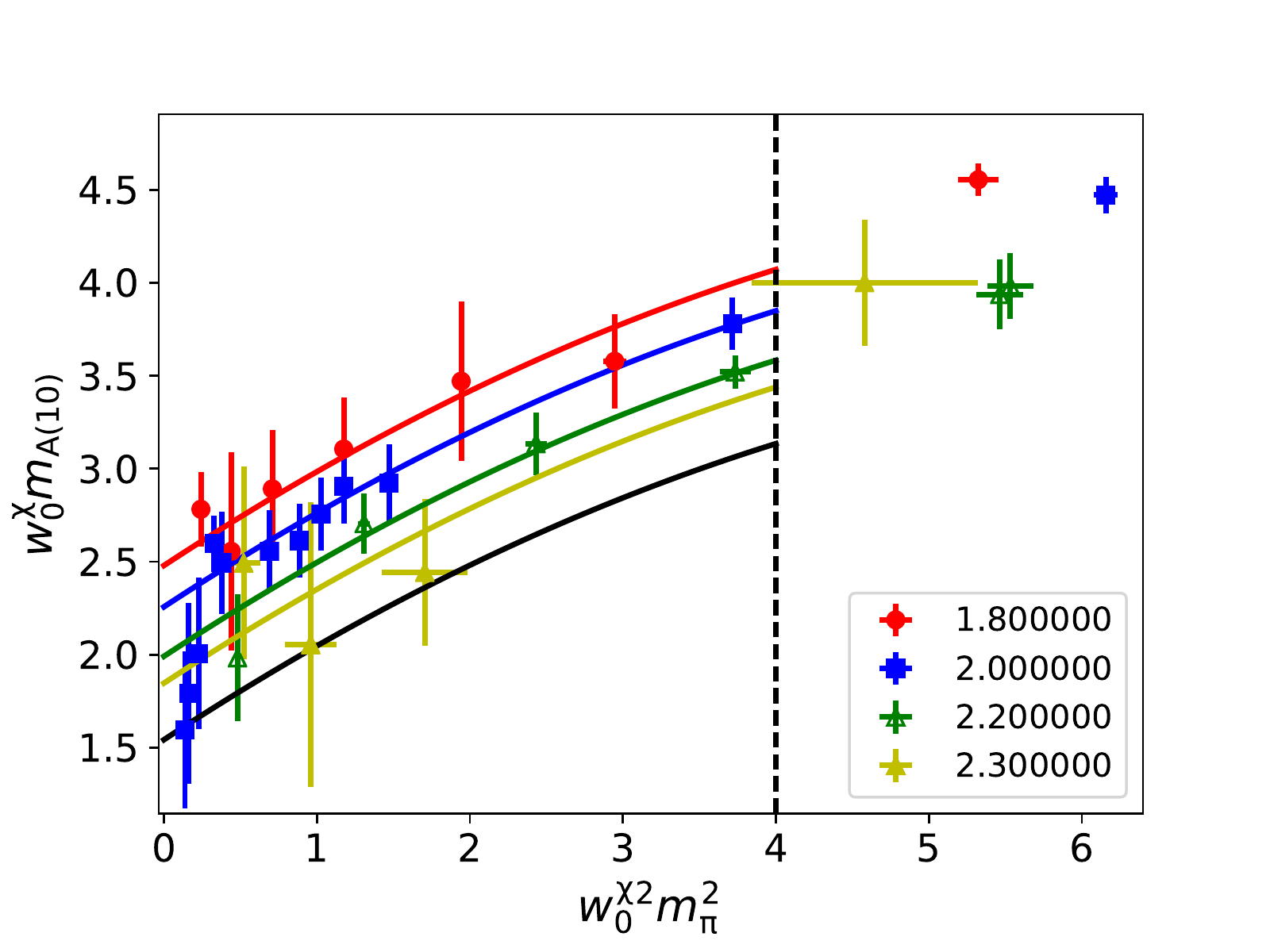}
  \end{center}
  \caption{Chiral and continuum extrapolation of the axial meson in 5-dimensional (left) and 10-dimensional (right) representations of flavour symmetry group.}
  \label{fig:ma}
\end{figure}
    $w_0^\chi m_{A5}^\chi = 1.09(15)$   with $ \chi^2/dof = 0.306   $ and 
    $w_0^\chi m_{A10}^\chi = 1.54(18)$   with  $\chi^2/dof = 0.500  $.
This seems to support our analysis in the previous section that there are two distinct axial vectors in different representations of the flavour group.

In the future, we are also planning to update the analysis of the flavour singlet spectrum as first described in \cite{Arthur:2016ozw,Drach:2017jsh}.

\section{Conclusions}
We presented updated results for the spectrum of SU(2) gauge theory with 2 quarks in the fundamental representation, which is the minimal realization of a UV-complete composite Goldstone Higgs model. 
The spectrum of the states of this model has been classified in terms of spin, flavour and parity, and we reported the interpolating operators associated with each of the states and possible decay products.
We have presented improved numerical results which were first published in \cite{Arthur:2016dir}. Our new estimates for the spectrum are consistent with the previous results.
We find that, somewhat unexpectedly, the vector meson mass has very small discretisation effects and is approximatively equal to 1 in units of $w_0^\chi$.
We also measure two distinct masses for the axial resonances in {\bf 5} and {\bf 10} representations of Sp(4). Our results for the low-energy constants and meson spectrum can be summarised in the following table:
  \begin{center}
    \begin{tabular}{ccccc}
      $w_0^\chi$ B & $w_0^\chi$ F & $w_0^\chi m_V$ &  $w_0^\chi m_A$ (5) & $w_0^\chi m_A$ (10) \\
      \hline
      3.01(11) & 0.0632(55) & 0.989(27) & 1.09(15) & 1.54(18) 
    \end{tabular}
  \end{center}

Future improvements will include improving the precision of renormalisation constants, which could lead to significant increase in the precision of the final results, and investigating the flavour singlet spectrum.
We are also working on investigating the structure of the vector resonance, an equivalent to $\rho\to\pi\pi$ decay in QCD, which would have observable effects in vector meson scattering.

This work was supported by the Danish National Research Foundation DNRF:90 grant and by a Lundbeck Foundation Fellowship grant. The computational resources were provided by the DeIC national HPC centre at SDU.
This work was supported by a grant from the Swiss National Supercomputing Centre (CSCS) under project ID s688.
\bibliography{lattice2017}

%%%%%%%%%%%%%%%%%%%%%%%%%%%%%%%%%%%%%%%%%%%%%%%%%%%%%%%%%%%%%%%%%%%%%%%%%%%%%
\end{document}